\documentclass[11pt]{article}
\newsavebox{\PSLASH}
\sbox{\PSLASH}{$p$\hspace{-1.8mm}/}

\begin{document}
\title{\large \bf A note on Pathria's model of the universe as a black hole}
\author{ S. Khakshournia
\footnote{Email address: skhakshour@aeoi.org.ir}
\\
\\
Nuclear Science and Technology Research Institute (NSTRI), Tehran, Iran\\
\\
\\
}\maketitle
\[\]

\begin{abstract}

Pathria has shown that for the certain values of the cosmological
constant, a pressureless closed Friedmann-Robertson-Walker
universe can be the interior of a Schwarzschild black hole. We
examine the Pathria's model from the point of view of the matching
of the two different spacetimes through a null hypersurface. We
first regard the event horizon of the black hole, which in the
Pathria's model is identified with the radius of the universe at
the point of its maximum expansion, as a null hypersurface
separating the FRW interior from the vacuum Schwarzschild
exterior, and then use the null shell formalism for it. It turns
out that the matching is not smooth, and in fact, the null
hypersurface is the history of a null shell admitting a surface
pressure.
\end{abstract}
\hspace*{1cm}PACS numbers: 98.80.-k; 04.20.Jb

\section{Introduction}
Pathria \cite{Path} published a paper showing that a pressureless
closed Friedmann-Robertson-Walker (FRW) universe can lie inside a
Schwarzschild black hole existing in a bigger universe (see also
\cite{Stuckey}). He has first considered a FRW universe with
positive 3D curvature, consisting of the dust matter, in a phase
of decelerating expansion, with the value of the cosmological
constant $\Lambda$ such that $0\leq\Lambda\leq \Lambda_{c}$, where
$\Lambda_{c}=(2\pi \rho R^3)^{-2}$, with $\rho(t)$ and $R(t)$ are
the energy density and the radius of curvature of the universe,
respectively. In this case, the solution for the function $R(t)$
shows that in a later epoch it reaches its maximum value $R_{max}$
increasing monotonically from the value $R_{max}=C$
($C=\frac{8\pi}{3}\rho R^{3})$ to $R_{max}=\frac{3}{2}C$, as
$\Lambda$ increases from $0$ to $\Lambda_{c}$. Pathria then
compared this behavior with that of a 4D Schwarzschild black hole
modified by the presence of the cosmological constant $\Lambda$
and noted that the solutions for the radius of the event horizon
$R_{s}$ of the black hole are identically the same as for
$R_{max}$, irrespective of the values of the parameters $C$ and
$\Lambda$, except that for $\Lambda > \Lambda_{c}$ neither $R_{s}$
nor $R_{max}$ exists \cite{Path}. He therefore suggested that the
identity $R_{max}\equiv R_{s}$ is fundamental to the structure of
the universe and concluded that at any epoch $R(t)\leq R_{s}$,
meaning that the universe is indeed inside a black hole. \\
We note that this identification implies that the separating
surface of the FRW interior from the vacuum exterior, while the
universe reaches its maximum expansion, turns out to be
(instantaneously) a null hypersurface. So what if we study the
Pathria's cosmological model from the matching conditions point of
view for the two different spacetimes through a null hypersurface?
Hence, in this note, we formulate the general problem of gluing a
FRW metric to a vacuum schwarzschild one along a null
hypersurface. For this purpose, we use Barrab\`{e}s-Israel null
shell formalism \cite{hogen} to investigate the matching conditions. \\
\textit{Conventions.} Natural geometrized units, in which $G=c=1$,
are used throughout the paper. The null hypersurface is denoted by
$\Sigma$. The symbol $|_{\Sigma}$ means "evaluated on the null
hypersurface ". We use square brackets [F] to denote the jump of
any quantity F across $\Sigma$. Latin indices range over the
intrinsic coordinates of $\Sigma$ denoted by $\xi^{a}$, and Greek
indices over the coordinates of the 4-manifolds.

\section{Matching Conditions}

We consider a pressureless closed FRW expanding universe which can
be described, in retarded time coordinates, by the following
metric
\begin{equation}\label{metricLTB}
ds^{2}_{-}=-a^2 dv(2d\chi+dv)+a^2\sin^2\chi(d\theta^{2} +\sin^{2}
\theta d\varphi^{2}),
\end{equation}
where $a(v,\chi)$ is the scale factor, $v$ is a null coordinate ,
and $\chi$ a spacelike coordinate which is constant along the
comoving worldlines. The retarded time $v$ is related to the
cosmological time $t_{-}$ by
\begin{equation}\label{advancedtime}
dv=\frac{dt_{-}}{a}-d\chi.
\end{equation}
The governing Friedmann equations are written as
\begin{equation}\label{par1}
\frac{\dot{a}^2}{a^2}+\frac{1}{a^2}-\frac{\Lambda}{3}=\frac{8\pi\rho}{3},
\end{equation}
\begin{equation}\label{par11}
\frac{\ddot{a}}{a}-\frac{\Lambda}{3}=-\frac{4\pi\rho}{3},
\end{equation}
where the dots denote the derivative  with respect to $t_{-}$;
$\Lambda$ and $\rho$ are the cosmological constant and the energy
density, respectively.\\
Considering an external view of the universe, we assume the
exterior of the FRW universe  may be described by a Schwarzschild
metric modified by the presence of the cosmological constant as
follows
\begin{equation}\label{metricext}
ds^{2}_{+}=-du(fdu+2dr)+r^{2}(d\theta^{2}+\sin^{2} \theta
d\varphi^{2}),
\end{equation}
where
\begin{equation}\label{fieldeq1}
f=1-\frac{2M}{r}-\frac{\Lambda r^2}{3},
\end{equation}
with the Schwarzschild mass $M=4\pi\int_{0}^{r}\rho
r^{2}dr=\frac{4\pi}{3}\rho r^{3}\big|_{\Sigma}$, being a constant
for a dust-filled universe.\\
 To match the  FRW interior with the Schwarzschild
exterior along the null hypersurface $\Sigma$ situated at the
maximum radius of the FRW expanding universe, we note that the
requirement of the continuity of the induced metric on $\Sigma$
yields the following matching conditions
\begin{eqnarray}\label{par2}
r&\stackrel{\Sigma}{=}&a\sin\chi, \nonumber\\
2d\chi&\stackrel{\Sigma}{=}&-dv,\\
fdu&\stackrel{\Sigma}{=}&-2dr,\nonumber
\end{eqnarray}
where $\stackrel{\Sigma}{=}$ means that we should compute both
sides of the equality at $\Sigma$. Taking $\xi
^{a}=(u,\theta,\varphi)$ with $a=1,2,3$ as the intrinsic
coordinates on $\Sigma$, and $u$ a parameter on the null
generators of the hypersurface, we now calculate the tangent basis
vectors $e_{a}=\partial/\partial \xi ^{a}$ on both sides of
$\Sigma$ as

\begin{eqnarray}\label{verbien}
e^{\mu}_{u}|_{-}&=&\left(\frac{du}{d\chi}\right)^{-1}\left(-2,1,0,0\right)\big|_{\Sigma},
\hspace*{1.4cm} e^{\mu}_{\theta}|_{-}=\delta^{\mu}_{\theta},
\hspace*{1.6cm}e^{\mu}_{\varphi}|_{-}=\delta^{\mu}_{\varphi},\\
e^{\mu}_{u}|_{+}&=&\left(1,-\frac{f}{2},0,0\right)\big|_{\Sigma},
\hspace*{2.5cm}e^{\mu}_{\theta}|_{+}=\delta^{\mu}_{\theta},
\hspace*{1.6cm}e^{\mu}_{\varphi}|_{+}=\delta^{\mu}_{\varphi}.
\end{eqnarray}
For further applications, using  Eq. (\ref{par2}) together with
(\ref{advancedtime}) and (\ref{par1}), we can compute
\begin{eqnarray}\label{trans2}
\frac{du}{d\chi}&=&\frac{du}{dr}\frac{dr}{d\chi},\nonumber\\
\hspace*{0.9cm}&=&\frac{-2}{f}\left(-\frac{\partial a}{\partial\chi}\sin\chi
+a\cos\chi\right)\big|_{\Sigma},\\
&=&\frac{-2}{f}\left(-a\dot{a}\sin\chi+a\cos\chi\right)\big|_{\Sigma}.\nonumber
\end{eqnarray}
Recalling that $u$ is a parameter on the null geodesic generators
of $\Sigma$, we choose the tangent-normal vector $n^{\mu}$
coinciding with the tangent basis vector associated with the
parameter $u$, so that $n^{\mu}=e^{\mu}_{u}$ \cite{Pois}. We may
then complete the basis by a transverse null vector $N^{\mu}$
uniquely defined by the four conditions $n_{\mu}N^{\mu}=-1$,
$N_{\mu}e^{\mu}_{A}=0$ $(A=\theta,\varphi)$, and
$N_{\mu}N^{\mu}=0$. We find
\begin{eqnarray}\label{normaltrans}
N_{\mu}|_{-}&=&\frac{1}{2}\left(\frac{du}{d\chi}\right)\left(1,0,0,0\right)\big|_{\Sigma},\\
N_{\mu}|_{+}&=&(-1,0,0,0)\big|_{\Sigma}.
\end{eqnarray}
Furthermore, the induced metric on $\Sigma$, given by $g_{ab} =
g_{\mu\nu}e^{\mu}_{a}e^{\nu}_{b}|_{\pm}$, is computed to be\\
$g_{ab}$ =diag$\left(0,r^{2},r^{2}\sin^{2}\theta\right)\big|_{\Sigma}$,
which is the same on both sides of the hypersurface. Defining a
pseudo-inverse of the induced metric $g_{ab}$ on $\Sigma$ as
$g_{*}^{ac}g_{bc} = \delta_{b}^{a}+ n^{a}N_{\mu}e^{\mu}_{b}$, with
$n^{a} = \delta^{a}_{u}$ \cite{Bar1}, one
gets $g^{ab}_{*}$ = diag$\left(0,\frac{1}{r^{2}},\frac{1}{r^{2}\sin^{2}\theta}\right)\big|_{\Sigma}$.\\
The final junction conditions are formulated in terms of the jump
in the transverse extrinsic curvature tensor across $\Sigma$.
Using the definition ${\cal
K}_{ab}=e^{\mu}_{a}e^{\nu}_{b}\nabla_{\mu}N_{\nu}$, we may
therefore compute the transverse extrinsic curvature \cite{Bar1}
on both sides of $\Sigma$. Its nonvanishing components on the
minus side are found as
\begin{eqnarray}\label{K33m} {\cal
K}_{\theta\theta}|_{-}&=&\sin^{-2}\theta {\cal
K}_{\varphi\varphi}|_{-}\\\nonumber
&=&a\sin\chi\left(\frac{\cos^2\chi-
\dot{a}^2\sin^2\chi}{f}\right)\big|_{\Sigma},\\\nonumber
&=&a\sin\chi \big|_{\Sigma},
\end{eqnarray}

\begin{eqnarray}\label{K11m}
{\cal K}_{uu}|_{-}&=&\frac{-1}{2}f_{,r}-2\pi \rho
a^3f\left(\frac{dr}{d\chi}\right)^{-2}\sin\chi\big|_{\Sigma},\\\nonumber
&=&\frac{-1}{2}f_{,r}+2\pi\rho a\big|_{\Sigma},
\end{eqnarray}
where we have used Eqs. (\ref{par1}), (\ref{par11}),
(\ref{fieldeq1}), (\ref{par2}) and (\ref{trans2}), and the term
$\frac{dr}{d\chi}$ in Eq. (\ref{K11m}) has been evaluated at
$\chi=\frac{\pi}{2}$, corresponding to the maximum expansion of
the FRW universe. The nonvanishing components of the transverse
curvature on the plus side are
\begin{equation}\label{Ktt}
{\cal K}_{\theta\theta}|_{+}=\sin^{-2}\theta {\cal
K}_{\varphi\varphi}|_{+}=r\big|_{\Sigma},
\end{equation}
\begin{equation}\label{Kuu}
{\cal K}_{uu}|_{+}=\frac{-1}{2}f_{,r}\big|_{\Sigma}.
\end{equation}
Now  Eqs.  (\ref{K33m}) and  (\ref{Ktt}) show that the angular
components of the  transverse curvature are continuous across
$\Sigma$. But from  Eqs. (\ref{K11m}) and (\ref{Kuu}) we infer the
discontinuity of the "uu" component of extrinsic curvature across
$\Sigma$, signaling the presence of a null shell of matter on the
hypersurface $\Sigma$, with vanishing energy density, but a
surface pressure which is computed as \cite{Bar1}
\begin{eqnarray}\label{nullpressure}
p&=&-\frac{1}{8\pi}n^{a}n^{b}[{\cal K}_{ab}],\nonumber\\
&=&-\frac{1}{8\pi}[{\cal K}_{uu}],\\
&=&\frac{\rho a}{4}\big|_{\Sigma}.\nonumber
\end{eqnarray}
The surface energy-momentum tensor of a lightlike shell is then
written as $S^{ab}= pg_{*}^{ab}$, where the surface pressure $p$
is measured by the observers  comoving  with the cosmological dust
fluid and intersecting the hypersurface. \\
Therefore, the null hypersurface $\Sigma$, separating the closed
FRW interior at the point of its maximum expansion from the vacuum
Schwarzschild exterior, is the history of a null shell
characterized by its surface pressure (\ref{nullpressure}).\\

\section{Conclusion}

Applying the Barrab\`{e}s-Israel null shell formalism, we have
studied Pathria's cosmological model in which a pressureless
closed FRW universe is entrapped within a Schwarzschild black
hole, from the point of view of the matching conditions for the
two different spacetimes glued together along a null hypersurface.
As a result of the matching conditions, we see that the transition
from the expansion to contraction in the FRW universe located on
the event horizon of the Schwarzschild black hole can only be done
through a null shell  which is simply characterized by a surface pressure. \\
\hspace*{0.3cm}\textit{Note added in the published version}: Very
recently, Knutsen \cite{knut} critically examined Pathria's model
from the point of view of smooth matching of a FRW interior region
to the vacuum Schwarzschild exterior spacetime along a timelike
hypersurface. He points out that with the ill-employed notations
in the Pathria's work, the matching conditions yield the
unexpected results suggesting that  Pathria's idea of the Universe
as a black hole is untenable. However, it is worth addressing that
in  the limit when the timelike hypersurface separating the two
spacetimes as studied in \cite{knut}, is moved to the
Schwarzschild horizon, it becomes a null hypersurface through
which we have investigated the matching of the two spacetimes in
the present work.
\subsection*{Acknowledgment}
The author would like to thank the referee for helpful comments.

\end{document}